# Comprehensive Movie Recommendation System

Hrisav Bhowmick
*Dept. of Data Science*
*Praxis Business School*
Kolkata, Country
hrisav.bhowmick@gmail.com

Ananda Chatterjee
*Dept. of Data Science*
*Praxis Business School*
Kolkata, India
ananda.chatterjee89@gmail.com

Jaydip Sen
*Dept. of Data Science*
*Praxis Business School*
Kolkata, India
jaydip.sen@acm.org

*Abstract*— A recommender system, also known as a recommendation system, is a type of information filtering system that attempts to forecast a user's rating or preference for an item. This article designs and implements a complete movie recommendation system prototype based on the Genre, Pearson Correlation Coefficient, Cosine Similarity, KNN-Based, Content-Based Filtering using TFIDF and SVD, Collaborative Filtering using TFIDF and SVD, Surprise Library based recommendation system technology. Apart from that in this paper, we present a novel idea that applies machine learning techniques to construct a cluster for the movie based on genres and then observes the inertia value number of clusters were defined. The constraints of the approaches discussed in this work have been described, as well as how one strategy overcomes the disadvantages of another. The whole work has been done on the dataset Movie Lens present at the group lens website which contains 100836 ratings and 3683 tag applications across 9742 movies. These data were created by 610 users between March 29, 1996, and September 24, 2018.

## I. INTRODUCTION

A recommendation system is a type of information filtering system that attempts to forecast a user's rating or preference for an item. Recommender systems, in simple terms, are algorithms that propose appropriate products to consumers. Depending on the industry, these relevant objects could be "movies to watch," "goods to buy," or "text to read. The recommendation system is of two types as Content-based Filtering and Collaborative Filtering. Item features are used in content-based filtering suggestions to recommend other items that are comparable to what the users enjoy based on their previous actions or feedback. It functions based on information gathered from the user, either explicitly (ratings) or implicitly (data) (clicking on a link). Collaborative Filtering can be split in two ways: user-based collaborative filtering, and item-based collaborative filtering. Similar users are found using user-based filtering. Similar users are those who share similar interests and tastes in movies, literature, and other media. The item-based filtering, on the other hand, finds similar items. In recommender systems, singular value decomposition (SVD) is employed as a method of collaborative filtering. A matrix structure is used in SVD with each row representing a user and each column representing an object. The ratings the user gives to the items are the constituents of the matrix. A content-based movie recommendation approach was documented by a few researchers [1]. An idea on collaborative filtering-based movie recommendation was also proposed by researchers [2]. A hybrid model-based movie recommendation system using K means clustering and genetic algorithm (GAs) to segment transformed userspace [3]. A comparison between the implementation of an item-based and user-based collaborative filtering on a map-reduce framework has been represented in the paper [4]. A method using visual features extracted from the picture like data posters and still frames has been presented [5]. Collaborative Filtering was classified as memory-based in memory-based and model-based recommendation and their applications and limitations were discussed in a paper [6]. The human emotions-based Movie Recommender System was described in a paper [7]. A hybridization of content and a collaborative based recommendation was proposed in a paper where weights were assigned in the content-based recommendations depending on user's importance and those weights are calculated using a set of linear regression equations derived from a social network graph that incorporates human judgments of item similarity [8]. Few researchers proposed an Emotion-based Movie Recommender System (E-MRS) which provides suggestions to users using a combination of collaborative and content-based filtering approaches. The suggestion is based on assumptions about a user's feelings and preferences, as well as the opinions of other users who are similar to them [9]. Researchers proposed a human-based movie recommendation system that observed a user while watching a portion of a movie, then analyzed the data that is their facial expressions and heart rate over time [10]. A hybrid recommender combining content and collaborative-based filtering technique was described to recommend restaurants [11]. A method was proposed in a study for estimating final ratings using a mixture of several ratings provided by various similarity measures. Our investigations show that this combination makes use of the variation within similarities and provides the target user with high-quality tailored suggestions. [12]. A highly scalable SVD based recommendation method was presented in a paper written by a few researchers [13]. An incremental algorithm based on SVD was proposed in a paper which combines incremental algorithm with the approximating SVD called the incremental appro SVD [14]. A paper was presented by a group of researchers that discussed the benefits of two approaches of recommendation: memory-based and model-based. Clusters were created using training data in that paper for data smoothing and neighborhood selection [15]. The Bayesian Co-Clustering method was introduced by a few researchers for movie recommendations where there are non-missing entries and to handle sparse matrices [16]. A weight-based similarity algorithm called IR-IUF++ for the recommendation was introduced in a paper [17]. A user-item relevance model was proposed by researchers and the application of log-based collaborative filtering was shown in their paper [18]. Different K-nearest neighbors (KNN) algorithm with different similarity metrics for recommendations was documented in the paper [19].

In this paper five methods have been incorporated for a recommendation, those are Collaborative Filtering, Content-based Filtering, Single value decomposition, Genre Based, Pearson Correlation coefficient based, Cosine Similarity, KNN (with cosine distant metric).

The paper has been organized into five sections. The problem statement which is intended to solve has been described in section II. A quick review of previous work done on the recommendations system has been documented in section III. The different methodologies adopted to provide recommendations have been provided in section IV. The experimental results of the methodologies adopted have been shown in section V.

## II. PROBLEM STATEMENT

With the proliferation of information, the ability to rapidly locate one's favorite film among a great number of options has become a critical issue. When the consumer does not have a certain movie in mind, a personalized suggestion system can be very useful. The methods which have been implemented in this paper are content-based filtering, collaborative filtering, singular value decomposition, genre-based recommendation, Pearson correlation coefficient method, Cosine Similarity method, KNN using cosine distant metric. A gamut of methods was explained in this paper and used for recommendations.

## III. RELATED WORK

The current state of work has been divided into eight methods. Those are Genre Based, Pearson Correlation Coefficient Based, Cosine Similarity-Based, Clustering of Movies Based, Content latent matrix using TFIDF & SVD, Collaborative latent matrix with TFIDF & SVD Based, Surprise Library (with KNN Basic) Based Recommendations. In the applied approaches in our study, most of them relied on content-based filtering. Content-based filtering, unlike collaborative filtering, does not rely on data from other users to make recommendations. A content-based filtering system can start producing relevant recommendations after a user has searched and browsed a few goods and/or completed certain transactions. The content-based strategy is used because it works well in domains or circumstances where the number of items exceeds the number of users. Related to the method used in our study a paper was released using genre-based recommendation on Movie Lens Dataset where content-based filtering technique was used using genre correlation and the data analysis tool which was used is R [21]. The Pearson Correlation Coefficient (PCC) is one of the most commonly used similarity measures in collaborative filtering recommender systems to determine how closely two users are linked. A paper was presented where an extension toward the Pearson correlation coefficient was used as a measure where the similarity between users doesn't exist [22]. A paper was released that used the TF-IDF (Term Frequency Inverse Document Frequency) and cosine similarity methods [23]. In that paper, a news recommendation system was designed to compute the cosine similarity of feature vectors obtained through the TF-IDF method, and the proposed method experimented on Slovakia newspapers. For a collaborative filtering-based recommender system, an Adaptive KNN, a new form of the KNN algorithm was suggested in a paper to address problems such as scalability, sparsity [24]. The suggested recommendation method is tested using the standard Movie Lens dataset, and the outcomes are assessed using Precision, Recall, F-Measure, and Accuracy. With the highly sparse data used for recommendation creation, the testing results showed that our suggested AKNN algorithm outperformed previous algorithms. In another work, the DBSCAN clustering method was used to develop a recommendation system. And then different voting algorithms were combined to recommend products to users based on which cluster they belong to [25]. The goal of the research was to reduce the running time of the algorithm while maintaining a high standard of recommendation. In another research, a novel matrix factorization model was introduced called Enhanced SVD (ESVD), which combines traditional matrix factorization methods with active learning-inspired rating completion. In addition, a link was constructed between prediction accuracy and matrix density to further investigate its potentials. offer Multi-layer ESVD was also offered, which iteratively learns the model to increase prediction accuracy. The Item-wise ESVD and User-wise ESVD were provided to handle imbalanced data sets with considerably more users than items or far more items than users. On the well-known Netflix and Movie Lens data sets, the proposed methodologies were tested [26]. The architecture, implementation, and primary features of the Surprise, a Python library for Recommender System, were explained in a paper [27]. In addition, a comparison of the Surprise framework to other analogous frameworks was offered to show why it is superior to others in terms of building and handling new and complicated recommender models. Using the real-world benchmark Movie Lens datasets, it was examined the accuracy of various built-in similarity measures given by the Surprise framework (100 k and 1 M).

## III. METHODOLOGY

Eight methods have been applied in this work for recommendations. Genre-based recommendation, Pearson Correlation coefficient-based recommendation, Cosine Similarity-based recommendation, KNN based recommendation with cosine as the metric, K-means clustering based recommendation, latent matrix using TFIDF & SVD based on content-based filtering, latent matrix using TFIDF & SVD based on collaborative filtering, Surprise Library (with KNN Basic) Based Recommendations. Among TFIDF & SVD based recommendations and the cosine similarity-based recommendation the cosine function was applied on the feature vectors produced using *TfidfVectorizer()* function yielded the best results.

The data has been extracted from the Movie Lens dataset which contains 100836 ratings and 3683 tag applications across 9742 movies. These data were created by 610 users between March 29, 1996, and September 24, 2018. Every method adopted in this study has been explained below. The movie dataset has 9742 records and 3 columns which are *movieId, title, genres*. The rating dataset contains 100836 records and 5 columns. The tags dataset is contained of 3683 records and 4 columns.

*A. Genre Based*

The genre-based recommendation falls under content-based filtering. This form of recommendation system displays relevant items based on the content of the users' previously searched items. The attribute/tag of the product that the user likes is referred to as content in this case. Items are labeled with keywords in this type of system, after which the system tries to comprehend what the user wants by searching its database, and lastly tries to recommend different products that the user wants.

In genre-based recommendation is based on which genre the user prefers to watch. Suppose if the 'action' genre is preferred to watch by a user then that user will be recommended top movies from the action genre based on weighted score. The formula of the weighted score is as below.

Score= (v/v+m) *R) + (m/(m+v) *C), where

v = number of ratings for the movie

m = minimum number of ratings required to be eligible

R = average rating of the movie

C = mean rating across whole data

The dataset was stored in a data frame named *df_movies* containing three columns *movieid, title, genres*. The genres were separated by ('|'), which were removed using the *split ()* function, and then the list of genres was stored in a column called *genrelist* column in the same data frame. Then using a *counter()* function on the *genrelist* column it was seen the number of the genre that appears for a particular movie. Suppose Jumanji (1995) fell into three genres Adventure, Animation, and Children, so for this movie, the count of each genre came out as 1 and that count was stored in a *variable genre*. Thereafter a data frame was formed named genres forming a matrix where the count of each genre was in the column of the matrix and the value of that matrix was the count of genres. NAN values in the said matrix were replaced with the *fillna()* function. Thereafter the genres data frame was joined with the *df_movies_new* data frame using the *join ()* function and the *genrelist* column was dropped. The value in the genres data frame was in float type, which was converted to an integer using the *pd.to_numeric()* function. Another data frame *df_rating* was created after reading dataset 'ratings.csv' which contains 3 columns *userId*, *movieId*, and rating. A data frame named movie_rating was created after merging the data frames *df_movies_new* and *df_rating*. Thereafter mean of rating was evaluated and arranged in descending order after grouping by title and stored in a data frame *df1*, and in the same way count of rating was found out after grouping by title and stored in a data frame named *df2*. Then these two data frames *df1*, and *df2* were merged and stored in a data frame named *df*. Thereafter mean rating across the whole data was determined from df data frame and stored in variable *C* as shown in eq. (1) and the minimum number of ratings was calculated as 7 and stored in variable m as shown in eq. (2). A data frame was created named eligible where those movies were stored which has a minimum 7 number of ratings. A function was created named weighted rating through which three parameters were passed such as the number of ratings, the minimum number of rating, and mean rating to get the score using the above formula shown in eq. (1). Then a column named score was created in the data frame *eligible* containing values of the score. Thereafter *eligible_movies_full* data frame was created after merging the data frames *df_movies_new*, and *eligible*. Then a function was created *genre_wise_movies* where when a particular genre will be passed first presence of that genre in a movie will be checked, if found then the title of those movies will be listed out based on the score arranged in descending order.

After applying the complete algorithm developed by us the top 5 movies from the 'action' genre were selected as Fight Club (1993), Star Wars: Episode IV- A New Hope (1997), Dark Knight (2008), Princess Bride, The (1987), and Star Wars: Episode V - The Empire Strikes Back

*B. Pearson Correlation Coefficient*

Pearson's Correlation Coefficient is a straightforward yet powerful method for determining how one variable changes linearly with another. The advantage of this method has been applied in this recommender system. Following is the formula to calculate the correlation coefficient.

$$r_{xy} = \frac{\sum_{i=1}^{n}(x_i - \bar{x})(y_i - \bar{y})}{\sqrt{\sum_{i=1}^{n}(x_i - \bar{x})^2}\sqrt{\sum_{i=1}^{n}(y_i - \bar{y})^2}}$$

Where $x_i$, $y_i$, are individual sample points, n is the sample size, $\bar{x}$ is the mean, and $r_{xy}$ is the symbol to define the correlation coefficient.

At first, a pivot table was formed based on the *movie_rating* data frame where the index of the table was movie titles, the columns were *user_id* and the values were ratings given by the user, i.e rating of a particular movie given by a particular user was given and if that movie was not rated by some user, then that field was represented by NAN. The pivot table was stored in a data frame *movie_matrix*. Thereafter using the *corrwith()* function, the correlation of the top five movies for a given movie was found out. The range of Pearson Correlation Coefficient values lie between -1 to +1, the more it tends to +1 better is the correlation of that movie with the other one. Following the above procedure, a 'Titanic (1997)' was passed through the movie matrix and the output was top five movies highly correlated with Titanic (1997) those are Gossip (2000), White Man's Burden (1995), Rapid Fire (1992), Imaginary Heroes (2004), Killer Elite (2011).

*C. Cosine Similarity*

This approach determines how similar two users are by calculating the cosine of the angle between two vectors, here vectors are nothing but the movies. The following formula calculates the cosine similarity between two movies.

$$\text{Sim(x,y)} = \cos(\vec{X}, \vec{Y}) = \frac{\vec{X}.\vec{Y}}{|X|.|Y|}$$

Here X and Y are the two movies between which the cosine angle is to be found out to determine how similar they are. More the angle tends to 0 more movies will be similar. To apply this method in our study firstly a column in the *eligible_movies_full* data frame was created named indexcol to store the index values of the eligible movies. Therefore using *TfidfVectorizer()* function genres of *eligible_movies_full* data frame were converted from text to feature vectors and stored in a variable named *count_matrix* and thereafter using applying *cosine_similarity()* function on *count_matrix* cosine similarity values of a feature vector with others were evaluated and stored in a variable named *cosine_sim*. To get the cosine similar movie names first index value of a movie was found out passing the movie title to the *indexcol* of *eligible_movies_full* data frame, and then passing that index value to *cosine_sim* variable the cosine similarity values of that movie with the rest of the movies were determined and those similarity values were kept in *similar_movies* variable Therefore the similarity values were sorted in descending order and the and using a lambda() function similarity values of a movie with the other movies in descending order was found out discarding the first element as it is the similarity with itself. Thereafter running a for loop on the similar_movies variable it was checked the first element of the similar_movies variable which is the index number of the movie is matched with which index of movies in eligible_full_movies data frame in the indexcol column and then the title of that movie was printed, in this way after iterating the loop for 5 times it was terminated using break() function to print top 5 movies name having cosine similar with the given movie.

In our case, Titanic (1997) was passed as the movie title and the top 5 movies which are cosine similar to this movie were obtained as Leaving Las Vegas (1995), Persuasion (1995), How to Make an American Quilt (1995), Bed of Roses (1996), and Angels and Insects (1995).

*D.    KNN algorithm( with cosine metric)*

KNN algorithm-based recommendation falls under collaborative filtering. In our work, item-based collaborative filtering has been used where the similarity between a particular item and the K other particular items have been calculated using Euclidean distance and cosine similarity. Here less the value of the cosine angle between the target item and other items more they are similar. The value of K has been taken as 5 in our case where the least distance between the top 5 movies with the target movie has been evaluated and then recommended to the user who watched the target movie.

Firstly, a pivot table has been generated from the *movie_rating* data frame where the movie title has been assigned to the index, the user id has been assigned to the column, and the ratings to the movie given by each user have been assigned as the value of the matrix. The pivot table has been stored in a data frame called *rating_matrix*. If a particular movie has not been rated by some user in that case that field has been filled up by NAN. Those NAN values were replaced with 0 using the *fillna()* function. Thereafter, the pivot table has been converted to an array matrix and stored in a data frame called *rating_matrix_new*. Thereafter *NearestNeighbors* class was imported from the *sklearn.neighbors* library and parameters metric, and algorithm were passed through it. The metric was assigned as 'cosine' to find out the similarity and the algorithm was assigned as 'brute' and the class was stored in a variable named *model_knn*. Thereafter *model_knn* was fitted on the *rating_matrix_new* data frame. After that, an index of any movie was selected. In our case, it was 5617 stored at variable *query_index*.and then to find out the movies similar to the movie at index value 5617 *model_knn.Kneigbours()* function was used and in that function that particular index was passed as *rating_matrix.iloc[query_index,:]* which will give out all the features of the movie stored at index 5617, also another parameter *n_neighbours* was passed and assigned value 6 to obtain 6 movies similar to the movie stored at that particular index. This function *model_knn.kneighbors()* gives out two features distances and indices. Thereafter using *flatten()* function the distance was flattened and for a particular movie stored in index 5617 the other recommended movies with distance came out and this process was done by running a for loop over the length of distance feature.

In our case, the movie stored in index 5617 came out as Mezzo Forte (1998) and the recommended movies situated at a shorter distance with this came out in the order of short distance such as Guyver: Dark Hero (1994), Harrison Bergeron (1995), Real Life (1979), The Punisher: Dirty Laundry (2012), House of Cards (1993).

*E.    Clustering of Movies*

Clustering is another approach of recommending movies where by looking at the movies a user watch clustering is used to find similar movies and ultimately propose the ones that are the most similar. K-means clustering algorithm has been used in our study for the recommendation. The K-means algorithm's main goal is to reduce the sum of distances between points and their corresponding cluster centroid. Here the concept of inertia comes into action which calculates the sum of the distances of all the movies within a cluster from their cluster centroid. The objective is to keep the inertia lower in a cluster and as far as possible in two different clusters.

To apply the K-means algorithm in our study first a for loop was run over 9 clusters and those 9 clusters were passed in a *KMeans ()* function and fitted on *count_matrix* data frame which contains the feature vectors of genres of movies. Thereafter an empty list named inertia was taken which was appended by the inertia value of each cluster and from looking at the values of inertia we decided to consider 6 clusters among 9 as after cluster 5 the distance between two different clusters came out very small which is violating the cluster forming criteria. Then clusters label was stored in a list form to a variable named cluster using km. *labels_watolist()* function wherein km is the instance of *kMeans(n_cluster=6)* function. Therefore another column named cluster was added to the *eligible_movies_full* data frame and stored in a data frame *eligible_movies_upd* Thereafter empty list was created for cluster0 to cluster5 and each of those lists were appended with the movie title falls under the respective cluster using the command if *eligible_movies_upd['cluster'].iloc[row]==0:* where *row* is the variable iterates over 6 clusters.

In our work following the above-described method the first 5 movies were obtained as Toy Story (1995), Jumanji (1995), Tom and Huck (1995), Balto (1995), Now and Then (1995).

*F.    Content latent matrix using TFIDF & SVD*

A dataset named tags.csv was taken and from their 3 columns named *userid*, *movieid*, and *tags* were chosen and stored in a data frame named *df_tag*. Thereafter *df_movies*, and df_tags data frames were merged using left join and stored in data frame movie_tags. Therefore two columns of *movie_tags* data frame named genres, and tag were joined using *join()* function passing through a lambda() function and kept in a column named metadata. After that text cleaning was done on the title column of *movie_tags* data frame using importing re module and cleaned text was stored in a column named title2. Therefore, the index of the *rating_matrix* data frame was stored in the *movie_list* variable and filtered the movies which are present in the *rating_matrix*, and stored the indexes of those movies in the column named *titleorder* in the *movie_tags_*new data frame. Thereafter feature vectors of the metadata column were formed using *Tfidf()* function on the metadata column and stored in a variable named *tfidf_matrix*. The values of *tfidf_matrix* were later converted to an array using the *toarray()* function and stored in a data frame named *tfidf_df*, where the rows were movie titles and vectorized forms as columns. The number of columns of that dataframe was 9658 which were compressed to 1000 applying SVD and to do that first *TruncatedSVD()* function with parameter N-components = 1000 were used and stored in a variable named svd and then that was fitted onto the *tfidf_df* data frame. A dataframe named *latent_matrix* was taken where all the movie indices along with those compressed 1000 columns containing user ids were stored and then using *cosine_similarity()* function on the *latent_matrix1* dataframe similar movies of a particular movie was figured out based on the closest cosine similarity score and for this purpose, the *latent_matrix1* data frame and *latent_matrix1.loc()* function was used through which a particular movie name was passed to find the similar type of movies.

In our case, using the above-explained method for the movie 'Batman Begins (2005)' the recommended movies that were found was 'Batman: Mystery of the Batwoman (2003)', Batman: Assault on Arkham (2014), Batman (1989), Batman: Year One (2011), Batman: The Killing Joke (2016).

G.   *Collaborative latent matrix with TFIDF & SVD*

In collaborative filtering user vs movie rating matrix was taken with userid in columns and movie title in rows. Thereafter applying SVD columns were compressed to 100 and for this purpose, the TruncatedSVD() function with parameter n_components=100 was fitted on rating_matrix dataframe and stored in the variable named latrent_matrix2. Therefore a data frame named latent_matrix2_df was formed with 100 columns. In the said data frame movie title was assigned as the index, userid was assigned as the column, and the ratings given by each user to each movie were assigned as the value of the rating_matrix. Therefore, the cosine similarity function was applied on latent_matrix2_df data frame and two parameters were passed through the function, one is the whole dataframe, and the second one is the particular movie name whose cosine similarity with other movies are about to be found out. Therefore top 5 movies were obtained based on the closest similarity values with the particular movie which was passed through the cosine similarity() function.

In our study, the particular movie selected as V for Vendetta (2006), and the top 5 recommended movies concerning this movie were Dark Knight, The (2008), Pirates of the Caribbean: The Curse of the Black Pearl (2003), Iron Man (2008), Kill Bill: Vol. 1 (2003), and 300 (2007).

H.   *Surprise Library (with KNN Basic)*

Initially, the data from the df_rating data frame consists of 3 columns such as userid, movieid, and rating. Thereafter KNNBasic() class was imported and stored in a variable named Knn. Thereafter 5 fold cross-validation was done using KNNBasic() and an RMSE score was obtained for each score. Therefore full data was taken as a training dataset and Knn() function was fitted on that. Thereafter for predicting the rating of the movieid 400 by the userid 26 was predicted as 3.5 and this prediction was done using the knn.predict() function. Here it was assumed that the userid 26 didn't rate the movieid 400.

For movie recommendations using KNNBasic, a function was developed named top 5movies through which 3 parameters were passed (userid, predictions, n=5) where n is the number of top 5 movies to be recommended. An empty dictionary named predict_ratings was taken, then a for loop was run and with each iteration, it was checked if the id matches with the actual userid then the dictionary at that particular movieid was appended with estimated ratings. Therefore the movieid was sorted in descending order based on the predicted rating. Thereafter the movieid was stored in a variable named top_movies after running the for loop over predicted_ratings. Therefore it was checked if the movieid stored in top_movies variable present in the movieid of df_movies data frame, if found then the corresponding movie title was printed.

In our case, for the userid 450, the movies were selected as top 5 were Braveheart (1995), Taxi Driver (1976), North by Northwest (1959), One Flew Over the Cuckoo's Nest (1975), Saving Private Ryan (1998).

IV.   EXPERIMENTAL RESULTS

The results obtained applying eight different described methods have been tabulated one by one below.

A.   *Genre Based Recommendation*

Table I.

| Title | Genres | Score |
|---|---|---|
| Fight Club (1999) | Action Crime Drama Thriller | 4.241499 |
| Star Wars: Episode IV - A New Hope (1977) | Action-Adventure Sci-Fi | 4.204796 |
| Dark Knight, The (2008) | Action Crime Drama IMAX | 4.194470 |
| Princess Bride, The (1987) | Action-Adventure Comedy Fantasy Romance | 4.186827 |
| Star Wars: Episode V The Empire Strikes Back | Action-Adventure Sci-Fi | 4.185033 |

Using the genre-based recommendation system the movies were recommended based on the score in descending order as shown above.

B.   *Pearson Correlation Coefficient Based Recommendation*

Table II.

| Title | Score |
|---|---|
| Gossip (2000) | 1.0 |
| White Man's Burden (1995) | 1.0 |
| Rapid Fire (1992) | 1.0 |
| Imaginary Heroes (2004) | 1.0 |
| Killer Elite (2011) | 1.0 |

The collaborative filtering approach was taken in this method wherein a pivot table was formed with the rows as the user id and the columns as the movie title. The correlation coefficients of a chosen movie with others were computed and based on that value movies were recommended to the target user in a descending order based on coefficient value as shown above. In the above table for the movie Titanic (1997) the above movies in an exact shown order were recommended.

*C. Cosine Similarity-Based Recommendation*

Table III.

| Title | Score |
|---|---|
| Leaving Las Vegas (1995) | 1.0 |
| Persuasion (1995) | 1.0 |
| How to Make an American Quilt (1995) | 1.0 |
| Bed of Roses (1996) | 1.0 |
| Angels and Insects (1995) | 1.0 |

TFIDF () function was applied to the genres of the movies to evaluate the feature vectors of the genres then using cosine () function on those feature vectors cosine similarity for a particular movie belongs to a particular genre the movies were recommended which were having closest cosine values of that movie. In the above table for the movie title Titanic (1997) was the above-tabulated movies were recommended in the shown order.

*D. KNN algorithm (with Cosine distance metric) Based Recommendations*

Table IV

| Title | Score |
|---|---|
| Guyver: Dark Hero (1994) | 0.3035 |
| Harrison Bergeron (1995) | 0.3638 |
| Real Life (1979) | 0.3824 |
| The Punisher: Dirty Laundry (2012) | 0.3860 |
| House of Cards (1993) | 0.3874 |

The collaborative filtering technique was applied here. A sparse matrix was formed with rows as the movie title and the columns as the user id. Then fixing the number of nearest neighbors as 5 and applying the KNN algorithm closest 5 movies for a particular movie watched by a user was figured out using cosine value with the watched one. Then the Euclidean distance with the watched movie and the other 5 nearest movies were computed and arranged in the above-shown order according to the distance in the lowest to the highest order. Here lower the distance greater is the similarity of the movie with the chosen one. In our work, for the movie Mezzo Forte (1998): the recommended movies in the shown order are tabulated.

*E. Clustering Based Recommendation*

Table V

| Title | Genres | Cluster |
|---|---|---|
| Toy Story (1995) | Adventure Animation Children Comedy Fantasy | 3 |
| Jumanji (1995) | Adventure Children Fantasy | 3 |
| Grumpier Old Men (1995) | Comedy Romance | 0 |
| Waiting to Exhale (1995) | Comedy Drama Romance | 0 |
| Father of the Bride Part II (1995) | Comedy | 5 |

Table VI

| Title |
|---|
| Grumpier Old Men (1995) |
| Waiting to Exhale (1995) |
| Sabrina (1995) |
| American President, The (1995) |
| Cutthroat Island (1995) |

Here movies were clustered based on the K-Means method. Thereafter, based on the inertia value the number of clusters was fixed. In our case, it was 6. Table, V shows how different movies with different genres fall under different clusters from 0 to 6, and table VI exhibits the list of movies that fall under cluster 0.

*F. Content latent matrix using TFIDF & SVD Based Recommendation*

Table VII

| Title | Score |
|---|---|
| Batman (1966) | 0.671195 |
| Batman: Mystery of the Batwoman (2003) | 0.637891 |
| Batman: Assault on Arkham (2014) | 0.608765 |
| Batman (1989) | 0.589502 |
| Batman: Year One (2011) | 0.576331 |

A 'metadata' column was made which will have a collection of genres, tags, movie names. Then *tfidf( )* function was applied on metadata column which formed features vectors of 'metadata' it will return a sparse matrix with movie name in rows and vectorized form as columns compress no. of columns to 1000 components with SVD use cosine similarity to find the similar type of movies to each of the movie and recommend based on closest score. In our work for the movie title Batman Begins (2005) the above movies were recommended in the shown order according to their score in descending order.

*G. Collaborative latent matrix using TFIDF & SVD Based Recommendation*

Table VIII

| Title | Score |
|---|---|
| V for Vendetta (2006) | 0.827567 |
| Dark Knight, The (2008)) | 0.813589 |
| Pirates of the Caribbean: The Curse of the Black Pearl (2003) | 0.804868 |
| Iron Man (2008) | 0.796666 |
| Kill Bill: Vol. 1 (2003) | 0.788545 |

A 'metadata' column was made which will have a collection of genres, tags, movie names. Then *tfidf( )* function was applied on the metadata column which formed features vectors of 'metadata' it will return a sparse matrix which is the user vs movie rating matrix with *userid* in columns and movie name in rows and the vectorized values as matrix value. The number of columns was compressed to 100 components with SVD. Then using cosine similarity function was used to find out the similar type of movies to each of the movies and recommend based on the closest score. In our work for the movie title Batman Begins (2005) the above movies were recommended in the shown order according to their score in descending order.

*H. Surprise Library (with KNN Basic) Based Recommendation*

Table IX

| Title |
|---|
| Braveheart (1995) |
| Taxi Driver (1976) |
| North by Northwest (1959) |
| One Flew Over the Cuckoo's Nest (1975) |
| Saving Private Ryan (1998) |

In this method *userid, movieid, and rating* were taken. Then using KNN Basic 5-fold cross-validation was run and the RMSE scores of each fold were obtained. In our study for userid 26 didn't provide any rating for movieid 400, using KNN basic predicted rating was found as 3.50.

The top 5 recommended movies for the userid 400 applying the above method were as shown in the above table in the shown order.

## V. DISCUSSIONS AND CONCLUSIONS

Eight different methods have been implemented in this study for recommending movies. Genre Based recommendation technique was a simple one where movies falling under the specific genre are checked first then based on the score as shown in eq(1) are recommended. But in genre-based there remains a high probability that the recommended movies are not desired by the target user as the recommendation is done based on only genres, not user profile similarity is checked before recommending. In the Pearson Correlation Coefficient Based recommended system the similarity between users can be easily found out but it's a long formula-based method, which causes a lot of computational time and memory. In the cosine similarity method, two users are labeled as cosine similar if both of them have watched and rated the same movies but the ratings given by these two users can be very different but the cosine similarity formula doesn't take this into account. In cluster-based recommendations, movies are clustered based on genres. If there are few movies with many genres in common then those movies fall into the same cluster. So, this cluster-based recommendation is not ideal to recommend any particular user. The content-based filtering with TFIDF & SVD is an efficient method as using SVD number of columns get compressed. Moreover, TFIDF is an efficient method to get the feature vectors based on those values cosine similar movies are found but here as the user doesn't get involved so this method isn't ideal for finding out similar users. The best method applied in this work is collaborative filtering with TFIDF & SVD, as here both user and movie get collaborated and based on the movie rated by the user, the user similarity is found out but it has also some disadvantages like many users don't rate movies which causes sparsity in the matrix. Moreover if a new it suffers from the cold start problem which is for a new user his profile remains empty as no movie hasn't been rated by the user by that time so the taste of the user remains unknown to the system causing difficulty arises while recommending the movie to that user. The use surprise library has also been done in recommending movies in one of the applied methods which is an inbuilt python library, using that with the help of Knn algorithm user rating can be predicted and based on the predicted rating movies can be recommended for a particular user but it's an ideal approach for user recommendation as it doesn't user movie rating matrix.

A compact movie recommendation system has been designed to show different approaches and their merits and drawbacks have also been explained here. Among all the approaches documented the collaborative filtering-based approach involving the method of TFIDF and SVD comes out to be the best.